\input{psfig.sty}
\documentstyle[12pt]{article}
\newfont{\feff}{cmti10}
\begin{document}

\title{Two-Dimensional Turbulence in the Inverse Cascade Range }

\author{ Victor Yakhot\\
Department of Aerospace and Mechanical Engineering\\
Boston University, 
Boston, MA 02215 }

\maketitle

${\bf Abstract}.$
\noindent
Numerical and physical experiments on the forced two-dimensional Navier-Stokes
equations show that transverse velocity differences are described by  
``normal'' Kolmogorov scaling $<(\Delta v)^{2n}>\propto r^{\frac{2n}{3}}$
and obey a gaussian statistics. Since the non-trivial scaling is a sign of a
strong non-linearity of the problem, these two results seem to contradict 
each other. A theory, explaining this result is presented in this paper.
Strong  time-dependence of the large-scale features
of the flow ($\overline{u^{2}}\propto t$) results 
in   decoupling of the large-scale dynamics from
statistically steady-state small- scale random processes. 
This time-dependence is
also a
reason for the localness of the pressure-gradient terms in the equations
governing the small-scale velocity difference PDF's. The derived 
self-consistent
expression 
for the pressure gradient contributions lead to
the conclusion 
that the small-scale
transverse velocity differences are governed by a linear Langevin-like 
equation,  strirred by a non-local,  universal,  solution-depending gaussian 
random force. This explains the experimentally 
observed gaussian statistics 
of transverse velocity differences and their Kolmogorov scaling..
The solution for  the 
PDF of longitudinal velocity differences is
based on a smallness of the energy flux in two-dimensional turbulence.
The theory makes a few quantitative predictions which can be tested 
experimentally.

\newpage

\section{Introduction}

Theoretical prediction of two inertial ranges, consequence of
  both energy and enstrophy
concervation laws by the two- dimensional Euler equations, was and still 
is 
one of the most remarkable achievements of statistical hydrodynamics 
[1]. A direct
and most important outcome  of these conservation laws is the fact that if
a fluid   is stirred by a random (or non-random) forcing,  acting on a scale 
$l_{f}=1/k_{f}$,  the produced  energy is spent on creation of the large-scale
 ($l>l_{f}$)
flow which cannot be dissipated in the limit of the large Reynlds number
$\nu\rightarrow 0$. This means that the dissipation terms are irrelevant
in the inverse cascade range. Since the dissipation contributions are one of
the most difficult obstacles on the road toward turbulence theory 
(see below), one can
hope that in two dimensions the situation is greatly simplified. This hope is
supported by recent numerical and physical experiments showing that as long as
the integral scale $L_{i}\propto t^{\frac{3}{2}}$ is  much smaller than the
size of the system, the velocity field at the scales $L_{i}>>l>>l_{f}$ is a
stationary close-to-gaussian process characterized by the structure functions 

\begin{equation}
S_{n}=\overline{(u(x+r)-u(x))^{n}}\equiv \overline{(\Delta u)^{n}}\propto
(Pr)^{\frac{n}{3}}
\end{equation}

\noindent where the pumping rate $P$ is defined below [2]-[4]. Moreover, both
numerical and physical experiments were not accurate enough to measure 

\begin{equation}
s_{2n+1}=\frac{S_{2n+1}}{S_{2}^{\frac{2n+1}{2}}}<<1
\end{equation}

\noindent which were too small. This means that the observed 
probability density
$P(\Delta u)$ was very close to symmetric one. This experimental fact 
 differs from the
outcome of the measurements  in three dimensions 
where $s_{n}$'s  are very large
when $n$ is not small. Thus,  the absence of strong (may be  any) 
 intermittency
in two-dimensional turbulence and proximity of the statistics of velocity
field to gaussian makes the problem seem tractable.

\noindent The equations of motion are (density $\rho\equiv 1$):

\begin{equation}
\partial_{t}v_{i}+v_{j}\partial_{j}v_{i}=-\partial_{i}p+\nu\nabla^{2}v_{i}+f_{i}
\end{equation}
\noindent and
\begin{equation}
\partial_{i}v_{i}=0
\end{equation}

\noindent where $\bf f$ is a forcing function mimicking the 
large-scale turbulence production mechanism and in a statistically 
steady state the mean pumping rate $P
=\overline{{\bf f\cdot v}}$. In the inverse cascade range the
dissipation terms in (3) will be irrelevant. Neglecting it and multiplying (3
) by $v_{i}$ we
obtain readily 

\begin{equation}
E=\frac{1}{2}\overline{v^{2}}=Pt
\end{equation}

\noindent Thus, in this case the energy linearly grows with time. 

In this paper we define the force correlation function as:

\begin{equation}
<f_{i}({\bf k})f_{j}({\bf k'})>\propto
P(\delta_{ij}-\frac{k_{i}k_{j}}{k^{2}})
\frac{\delta(k-k_{f})}{k}\delta({\bf k+k'})\delta(t-t')
\end{equation}

\noindent so that 

\begin{equation}
\overline{(f(x+r)-f(r))^{2}}\propto P(1-Cos(k_{f}r))
\end{equation}
\noindent

\noindent It will be clear below that the forcing term enters the equations
for the probability density of velocity differences  exclusively through the 
expression  (7) and in the limit $k_{f}r<<1$ it contribution is
$O((k_{f}r)^{2})$
which is a well-known fact. In the energy range we are interested in this work 
$k_{f}r>>1$ and the oscillating contribution can be neglected leading to
disappearence of the forcing scale from equation for the PDF.  Thus the general
expression for the structure functions is:

\begin{equation}
S_{n}(r)\propto (Pr)^{\frac{n}{3}}(\frac{r}{L_{i}(t)})^{\delta_{n}}
\end{equation}

\noindent where the exponents $\delta_{n}$ denote possible deviations from the
Kolmogorov scaling. If a statistically steady state exist in the limit
$L_{i}>>l>>l_{f}$,  then all $\delta_{n}=0$ since $L_{i}\propto
t^{\frac{3}{2}}$. This would be  prooof of `` normal'' (Kolmogorov) scaling 
in the inverse cascade range, provided one can show that the PDF $P(\Delta u)$
in the inertial range is independent on its counterpart in the interval
$l\approx l_{f}$. This is the subject of the present paper which is organized
as follows. In the next Section the equations for the generating functions 
are introduced. Section 3 is devoted to a short analysis of the Polyakov
theory of Burgers turbulence  some aspects of which are used in this paper. 
Some physical considerations, which are basis for the developing theory,
  are presented in Section 4. S
In Sections 5 and 6 the equations for the transvers and longitudinal
probability density functions are derived and solved. Summary and discussion
are presented in Section 7.

Now we would like to recall some well-known properties of velocity
correlation functions in incompressible fluids, needed below.
Consider two points
${\bf x}$ and ${\bf x'}$ and define ${\bf r}={\bf x-x'}$. Assuming that
the $x$-axis is paralel to the displacement vector ${\bf r}$,  one
can find that in the two-dimensional flow $d=2$ 
for the separation $r$ in the 
inertial range [5]-[7]:

\begin{equation}
\frac{1}{r^{d+1}}\partial_{r}r^{d+1}S_{3}=\frac{12}{d}{\cal E}
\end{equation}

\noindent giving  
\begin{equation}
S_{3}=\overline{(\Delta u)^{3}}
\equiv\overline{(u(x')-u(x))^{3}}\approx \frac{12}{d(d+2)}r
\end{equation}
\noindent and
\begin{equation}
S^{t}_{3}=\overline{(\Delta v)^{3}}
\equiv\overline{(v(x')-v(x))^{3}}\approx 0
\end{equation}

\noindent where $u$ and $v$ are the components of  velocity field paralel
and perpendicular to the $x$-axis  (vector ${\bf r})$. The relations (9)-(11)
resulting  from equations of motion (3) are dynamic properties of the velocity
field. Kinematics also gives something interesting:

\begin{equation}
\frac{1}{r^{d-1}}\frac{d}{dr}r^{d-1}S_{2}=(d-1)S_{2}^{t}\equiv\overline{(\Delta
v)^{2}}
\end{equation}

\noindent and in two dimensions we have:

\begin{equation}
S_{3t}\equiv \overline{\Delta u (\Delta v)^{2}}=\frac{1}{3}\frac{d}{dr}S_{3}
\end{equation}

\section{Equation for Generating Function}

\noindent  
We consider the $N$-point generating function:
\begin{equation}
Z=<e^{\lambda_{i}\cdot {\bf v(x_{i})}}>
\end{equation}

\noindent where the vectors ${\bf x_{i}}$ define  the positions of the points 
denoted $1\leq i \leq N$. 
Using the incompressibility condition,
the equation for $Z$ can be  written:
\begin{equation}
\frac{\partial Z}{\partial t}+\frac{\partial^{2} Z
}{\partial \lambda_{i,\mu}\partial  x_{i,\mu}}=I_{f}+I_{p}
\end{equation}

\noindent with 
\begin{equation}
I_{f}=\sum_{j} <{\bf \lambda_{j}\cdot f(x_{j})}e^{\lambda_{i}u(x_{i})}>
\end{equation}
\begin{equation}
I_{p}=-\sum_{j}\lambda_{j}<e^{\lambda_{i}u(x_{i})}\frac{\partial p(x_{j})}{\partial x_{j}}>
\end{equation}
\noindent The dissipation contributions have been neglected here as
irrelevant.

In what follows we will be mainly interested 
in the probability density function of the two-point velocity
differences which is ontained from (7)-(10), setting 
$\bf{\lambda_{1}+\lambda_{2}}=0$ (see Ref. [8] and the theory developed 
below), 
so that 
\begin{equation}
Z=<exp{(\bf{\lambda\cdot U})}>
\end{equation}

\noindent where 

\begin{equation}
{\bf U}={\bf u(x')-u(x)}\equiv \Delta {\bf u}
\end{equation}

\noindent 
The moments of the two-point velocity differences which
in 
homogeneous and isotropic turbulence can depend only on 
 the absolute values of two vectors
(velocity difference ${\bf v(x')-v(x)}$ and displacement 
${\bf r\equiv x'-x}$) and the angle $\theta$ between them with $\theta=\pi/2$
and $\theta=0$ corresponding to transverse and longitudinal structure 
functions, respectively. 
\noindent 
It is easy to show [5]- [6] that the 
 general form of the second-order
structure function in the inertial range is:
\begin{equation}
S_{2}(r,\theta)= \frac{2+\xi_{2}}{2}D_{LL}(r)(1-\frac{\xi_{2}}{2+\xi_{2}}cos^{2}(\theta))
\end{equation}
\noindent with $D_{LL}(r)=<(u(x)-u(x+r))^{2}>$.
 More involved relation can
 be written for the fourth-order moment:
\begin{equation}
S_{4}(r,\theta)=D_{LLLL}(r)cos^{4}(\theta)-3D_{LLNN}(r)sin^{2}(2\theta)+
D_{NNNN}(r)sin^{2}(\theta)
\end{equation}
\noindent where $D_{LLNN}=<(v(x)-v(x+r))^{2}(u(x)-u(x+r))^{2}>$
and $v$ and $u$ are the components of the velocity field perpendicular
and parallel to the $x$-axis, respectively. In general, 
in the llimit $cos(\theta)\equiv s\rightarrow \pm 1$, corresponding to the moments of the 
longitudinal velocity differences
$S_{n}(r,s)\rightarrow S_{n}(r)cos^{n}(\theta)$. 
This means that in this limit 
$Z(\lambda,r,s)\rightarrow Z(\lambda s,r)\equiv Z(\lambda_{x},r)$.
The generating function can depend only on three variables:

$$\eta_{1}=r;~~ \eta_{2}=\frac{{\bf \lambda\cdot r}}{{\bf r}}\equiv 
\lambda cos(\theta);~~ \eta_{3}=\sqrt{\lambda^{2}-\eta_{2}^{2}};$$  

In these variables:

\begin{equation}
Z_{t}+[\partial_{\eta_{1}}\partial_{\eta_{2}}+\frac{d-1}{r}\partial_{\eta_{2}}
+\frac{\eta_{3}}{r}\partial_{\eta_{2}}\partial_{\eta_{3}}+\frac{(2-d)\eta_{2}}{r\eta_{3}}\partial_{\eta_{3}}-\frac{\eta_{2}}{r}\partial^{2}_{\eta_{3}}]Z=
I_{f}+I_{p}
\end{equation}

\noindent where 
\begin{equation}
I_{p}=
\lambda_{i}<(\partial_{2,i} p(2)-\partial_{1,i} p(1))e^{\bf \lambda\cdot U}>
\end{equation}
\noindent and
\begin{equation}
I_{f}=(\eta_{2}^{2}+\eta_{3}^{2})P(1-Cos(k_{f}r))Z
\end{equation}

\noindent where, to simplify notation we set $\partial_{i,\alpha}\equiv
\frac{\partial}{\partial x.\alpha}$ and $v(i)\equiv v({\bf x_{i}})$.

\noindent In two dimensions 
the equation for the generating function becomes with $P=1$
(the subscript $o$ is omitted hereafter):

\begin{equation}
[\partial_{\eta_{1}}\partial_{\eta_{2}}+\frac{1}{r}\partial_{\eta_{2}}+
\frac{\eta_{3}}{r}\frac{\partial^{2}}{\partial_{\eta_{2}}\partial{\eta_{3}}}
-\frac{\eta_{2}}{r}\frac{\partial^{2}}{\partial \eta{_3}^{2}}-
(\eta_{2}^{2}+\eta_{3}^{2})]Z=I_{p}
\end{equation}

\noindent The generating function can be written as:

\begin{equation}
Z=<e^{\eta_{2}\Delta u + \eta_{3}\Delta v}>
\end{equation}

\noindent so that any correlation function 

\begin{equation}
<(\Delta u)^{n}\Delta v)^{m}>=\frac{\partial^{n}}{\partial
\eta_{2}^{n}}\frac{\partial^{n}}{\partial \eta_{3}^{m}}Z(\eta_{2}=\eta_{3}=0)
\end{equation}

\noindent Neglecting the pressure term $I_{p}$ and differentiating (25) once 
over $\eta_{2}$ we obtain immediately 

\begin{equation}
\frac{1}{r}\frac{d}{dr}rS_{2}=S_{2}^{t}
\end{equation}

\noindent Second differentiation (again neglecting $I_{p}$)  gives:

\begin{equation}
\frac{1}{r}\frac{d}{dr}rS_{3}-\frac{2}{r}S_{3t}-2=0
\end{equation}

\noindent Combined with (13) this expression gives

\begin{equation}
\frac{1}{r^{3}}\frac{d}{dr}r^{3}S_{3}-6=0
\end{equation}

\noindent which is nothing but the Kolmogorov relation, derived in 2d  without
contributions from the pressure terms. It follows from (25) that it is 
reasonable
to look for a scaling solution $Z(\eta_{2},\eta_{3},r)=Z(X_{2},X_{3})$ where
$X_{i}=\eta_{i}r^{\frac{1}{3}}$. 

\section{Polyakov's theory of Burgers turbulence}

The dissipation-generated 
contributions $O(\nu \nabla^{2} \overline{u_{i}u_{j}})\neq 0$ in the limit
$\nu\rightarrow 0$. This is a consequence of the ultra-violet singularity
$\nabla^{2}\overline{u_{i}(x)u_{j}(x+r)}\rightarrow \infty$ when $r\rightarrow 0$ making the
theory (the closure problem) extremely difficult. 
The expression for this    ``dissipation anomaly'', part of  the equation
 for
the generating function,  was developed by Polyakov for
the problem of the one-dimensional Burgers equation stirred by the random
force [8]. Theory of two-dimensional turbulence is free from the troubles 
coming from the
ultra-violet (dissipation) singularities. Still, here  we review some of 
the aspects of 
Polyakov's theory which 
we believe are
of general interest and which will be most helpful below.
Polyakov considered a one-dimensional problem [8]:

\begin{equation}
u_{t}+uu_{x}=f+\nu u_{xx}
\end{equation}

\noindent where the random force is defined by the correlation function

\begin{equation}
\overline{f(x,t)f(x+r,t')}=\kappa(r)\delta(t-t')
\end{equation}

\noindent The equation for generating function, analogous to (14), is written
readily:

\begin{equation}
Z_{t}+\lambda_{j}\frac{\partial}{\partial
\lambda_{j}}\frac{1}{\lambda_{j}}\frac{\partial Z}{\partial r}=
\kappa(r_{ij})\lambda_{i}\lambda_{j}Z+D
\end{equation}

\noindent where 
\begin{equation}
D=\nu \lambda_{j}<u''(x_{j},t)e^{\lambda_{k}u(x_{k},t)}>
\end{equation}

\noindent In the limit $r_{ij}\rightarrow 0$ the force correlation function 
$\kappa(r_{ij})=O(1-r_{ij}^{2})$ which imposes scaling properties of the
velocity correlation functions. In general,  the generating function depends 
on both velocity differences $U_{-}=\Delta u=u(x_{i})-u(x_{j})$ and sums 
$U_{+}=u(x_{i})+u_(x_{j})$ which makes the problem very difficult.
Defining Galilean invariance as independence of the correlation functions on 
``non-universal'' single-point $u_{rms}^{2}=\overline{u^{2}}$, 
 Polyakov assumed that if all $|U_{-}|<<u_{rms}$ then $U_{-}$ and $U_{+}$ are
statistically independent and $\sum \lambda_{i}=0$. 
In this case (see (8)), introducing
$\mu=\lambda_{2}-\lambda_{1}$ and the two-point generating function 

\begin{equation}
Z(\mu)=<e^{\mu \Delta u}>
\end{equation}

\noindent the equation for $Z$ reads in a steady
state:

\begin{equation}
(\frac{\partial}{\partial \mu}-\frac{1}{\mu})\frac{\partial }{\partial r}Z=
-r^{2}\mu^{2}Z+D
\end{equation}

\noindent where 

\begin{equation}
D=\mu \nu<(u''(x+r)-u''(x))e^{\mu\Delta u}>
\end{equation}

\noindent It is clear that the $O(r^2)$ forcing term imposes the scaling
variable $\xi=\mu r$ and $Z=F(\mu r)$ where $F$ is a solution of the following equation:

\begin{equation}
 \xi F''-F'+\xi^{2}F=D
\end{equation}

\noindent The problem is in evaluation of the dissipation
contribution $D$. 

On the first glance one can attempt to neglect $D$ and solve the resulting
equation. This is not so simple, however. The Laplace transform of (38) gives an
equation for the probability density $P=\frac{1}{r}\Phi(\frac{U}{r})
\equiv \frac{1}{r}\Phi(X)$

$$\Phi''+X^{2}\Phi'+X\Phi=0$$

\noindent Introducing 

\begin{equation}
\Phi=Exp(-\frac{X^{3}}{6})\Psi
\end{equation}

\noindent gives

\begin{equation}
\Psi''=(\frac{X^{2}}{4}-2X)\Psi
\end{equation}

\noindent which is the Schrodinger equation for a particle in a potential
$U(X)=X^{4}/4-2X$ not having any positive solutions.

\noindent The positivity of
the probability density is a severe constraint on a possible solution of the
equation of motion. That is where the dissipation contribution $D$ comes to
the rescue. Polyakov proposed a self-consistent conjecture about the structure
of the dissipation term

\begin{equation}
D=(\frac{b}{\mu}+a)Z
\end{equation}

\noindent modifying the potential in the Schrodinger equation with the
coefficients $b$ and $a$ 
chosen to produce the zero-energy ground state corresponding to positive PDF.
According to Ref.(8) this expression is the only one satisfying the galileo
invariance of the small-scale dynamics. 

\noindent The fact that the one or multi-dimensional advection 
contributions to the equation for the generating function do not lead to
positive solutions for the PDF is a general phenomenon (see below). The
importance of Polyakov's theory is, among other things,
 in realization that the dynamic closures for the remaining terms
must remove this problem. This
dramatically narrows the allowed classes of  closures. Thus, the
expressions for, $D$ or the pressure terms (see below), combined with
advective contributions to equation for $Z$ can be correct only and only if 
they lead 
to positive  
solutions for the PDF's in the entire range where $|\Delta u|<<u_{rms}$ and $r<<L_{i}$.

\section{Physical Considerations}
The problem of two-dimensional turbulence is simlified by the fact that the
dissipation contributions are irrelevant on the scales $l>>l_{f}$ we are
interested in. Moreover, since $u_{rms}$ grows with time, 
the statistically steady small-scale velocity differences 
$U_{-}=\Delta u$  with $r<<L(t)$ must be
decoupled from $U_{+}$ in (25). This means that the terms 

\begin{equation}
\overline{(\Delta u)^{n}(\Delta v)^{m}}
\end{equation}

\noindent can eneter the equation for $P(\Delta u,r)$ while the ones, 
involving

\begin{equation}
\overline{(\Delta u)^{n}(\Delta v)^{m}U_{+}^{p}}
\end{equation}

\noindent cannot. In principle, it can happen that the
$U_{-}U_{+}$-correlation functions can sum up into something time-independent.
However, at present we discard this bizarre possibility.

\noindent Next, the pressure gradients

\begin{equation}
\nabla p(x+r)-\nabla p(x)
\end{equation}

\noindent appearing in the equation (22)-(24) for $Z$ 
involve   integrals
over entire
space. It is clear that, if the steady state exists, 
 the large- scale contribution to the pressure integrals, 
depending on $L=L(t)$ cannot 
contribute to the small-scale steady-state dynamics, described by (25). 
That is why  the pressure contributions to $I_{p}$ (23) 
must depend  exclusively on the local scale $r$. This leads us to
an assumption that the pressure gradients in (23) are local 
in a sense the they can be expressed in terms of the velocity field at the
points $x$ and $x+r$. The application of these
considerations are presented below.

The theory of Burgers turbulence, dealt with the ``universal'' part of 
the dynamics, i.e. with the moments of velocity difference $S_{n}$ with
$n<1$. The theory of
two-dimensional turbulence, we are interested in, must produce the
moments with $n<\infty$ and that is why the algebtaic expressions for the
PDF's, characteristic of Burgers dynamics, are irrelevant. In addition, we
expect the small-scale dynamics in 2d to be independent on the forcing
function. This makes this problem very different.

\section{Transverse Structure Functions}

Unlike the probability density function for the longitudinal 
velocity differences
$P(\Delta u,r)$,
the transverse velocity difference  probability density 
is symmetric, i.e. $P(\Delta v,r)=P(-\Delta v,r)$. 
We are interested in the equation (25) in the limit $\eta_{2}\rightarrow 0$.
Let us first discuss some of the general properties of incompressible 
turbulence. Consider the forcing fucntion

$${\bf f}(x,y)=(f_{x}(x,y),0)$$

In this case the equation (25) is:

\begin{equation}
[\partial_{\eta_{1}}\partial_{\eta_{2}}+\frac{1}{r}\partial_{\eta_{2}}+
\frac{\eta_{3}}{r}\frac{\partial^{2}}{\partial_{\eta_{2}}\partial{\eta_{3}}}
-\frac{\eta_{2}}{r}\frac{\partial^{2}}{\partial \eta{_3}^{2}}-
\eta_{2}^{2}]Z=I_{p}
\end{equation}

\noindent Then, setting $\eta_{2}=0$ removes all  
information about forcing
function from the equation of motion. Based on our general intuition and
numerical data  we know that  two flows strirred by a one-component
or by a  two-component (statistically isotropic) forcing function  are
identical at the scales $l>>l_{f}$, provided the total fluxes 
generated by these
forcing functions are equal. This happens  due to  pressure terms

$$\Delta p=-\nabla_{i}\nabla_{j}v_{i}v_{j}$$

\noindent effectively mixing various components of the velocity field. This
universality, i.e. independence of the small-scale turbulence on the
symmetries  of
the  forcing, enables us to write an expression for the $I_{p}$
contribution to (25).

According to considerations, presented in a previous section, the pressure
gradients in the equation (25) are local and their dynamic 
role is in mixing various components of velocity
field. Thus the only contribution to $I_{p}$, not vanishing in the limit
$\eta_{2}\rightarrow 0$,  can be estimated as:

\begin{equation}
b\frac{\eta_{3}}{r}<\Delta u \Delta v e^{\eta_{2}\Delta u+\eta_{3}\Delta v}>=
b\frac{\eta_{3}}{r}
\frac{\partial}{\partial \eta_{2}}
<\Delta v e^{\eta_{2}\Delta u+\eta_{3}\Delta v}>
\end{equation}

\noindent Using  a theorem (see Frisch (8), for example) 
that for the random gaussian
process $\xi$ (see below) 

\begin{equation}
<\xi F(\xi)>=\overline{\xi^{2}}<\frac{\partial F(\xi)}{\partial \xi}>
\end{equation}

\noindent we derive in the limit $\eta_{2}\rightarrow 0$

\begin{equation}
I_{p}\approx b\eta_{3}^{2}\frac{\overline{(\Delta v)^{2}}}{r}\frac{\partial
Z_{3}}{\partial \eta_{2}}
\end{equation}

\noindent 
Substituting this into (25) and integrating over $\eta_{2}$ gives in the limit
$\eta_{2}\rightarrow 0$:

\begin{equation}
\frac{\partial Z_{3}}{\partial r}+\frac{Z_{3}}{r}+\frac{\eta_{3}}{r}\frac{\partial Z_{3}}{\partial
\eta_{3}}-\frac{\gamma}{r^{\frac{1}{3}}}\eta_{3}^{2}Z+
\Omega(\eta_{3})=\Gamma(\eta_{3})
\end{equation}

\noindent where $\gamma$ is undetermined parameter and an arbitrary function 

$$\Gamma(\eta_{3})=Z_{3}/r+\Omega(\eta_{3})$$

\noindent with 

$$-\Omega(\eta_{3})=
lim_{\eta_{2}\rightarrow 0}~\eta_{3}^{2}\int Z(\eta_{2},\eta_{3},r)d\eta_{2}$$ 

\noindent is chosen to satisfy 
a  trivial constraint
$Z_{3}(\eta_{3}=0,r)=1$ and the above mentioned universality.

\noindent 
This gives:

\begin{equation}
\frac{\partial Z_{3}}{\partial r}+\frac{\eta_{3}}{r}\frac{\partial Z_{3}}{\partial
\eta_{3}}-\frac{\gamma}{r^{\frac{1}{3}}}\eta_{3}^{2}Z=0
\end{equation}

\noindent where $Z_{3}=Z(\eta_{2}=0,\eta_{3})$. 
This equation is invariant under 
$\eta_{3}\rightarrow -\eta_{3}$ -  transformation. It is important that the
$O(\eta_{3}^{2})$ contribution to (50) comes from the pressure term but not
from the forcing, present in the original equation (25).
 Seeking a  solution 
to this
equation in a scaling form 
$Z_{3}(\eta_{3},r)=Z(\eta_{3}r^{\frac{1}{3}})\equiv Z(X)$ gives:

\begin{equation}
\frac{4X}{3}Z_{X}=\gamma X^{2}Z
\end{equation}

\noindent and 

\begin{equation}
Z=Exp(\frac{3\gamma}{8}\eta_{3}^{2}r^{\frac{2}{3}})
\end{equation}

This generating function corresponds to the gaussian distribution of
transverse velocity differences $P(\Delta v)$ with the second-order structure
function

\begin{equation}
S_{2}^{t}(r)=\overline{(\Delta v)^{2}}=\frac{3\gamma}{4}r^{\frac{2}{3}}
\end{equation}

The equation (50) correseponds to a  one-dimensional 
linear Langevin equation for ``velocity field'' $V=v/(Pr^){\frac{1}{3}}$

\begin{equation}
v_{\tau}(x)=-v(x)+\phi(x,\tau)
\end{equation}

\noindent where $\tau\propto tr^{-\frac{2}{3}}P^{\frac{1}{3}}$ and 
the non-local gaussian ``universal'' forcing $\phi(x,\tau)$,  
generated by the
nonlinearity of the original equation is defined by the correlation function

\begin{equation}
\overline{\phi(k,\tau)\phi(k',\tau')}\propto
\delta(k+k')\delta(\tau-\tau')
\end{equation}

\noindent The generating function for the field $V$ is 

$$z=<e^{XV}>$$

\noindent Since $\tau\propto tr^{-\frac{2}{3}}$ and $V\propto
vr^{-\frac{1}{3}}$
this equation is strongly non-local. It becomes local, however in the
wave-number space. This will be discussed later.

\noindent Now we can attempt to justify the relation (46). According to (23)
and taking into account that the $x$-axis is paralel to the displacement $r$ 
in the limit $\eta_{2}\rightarrow 0$

$$I_{p}\approx \eta_{3}<(\partial_{y}p(0)-\partial_{y'} p(r))
Exp(\eta_{3}\Delta v+\eta_{2}\Delta u)>$$

\noindent where
$$\partial_{y}p(0)-\partial_{y'} p(r)=\int
k_{y}(1-e^{ik_{x}r})[\frac{k_{x}^{2}}{k^{2}}u(q)u(k-q)+\frac{k_{y}^{2}}{k^{2}}v(q)v(k-q)+\frac{k_{x}k_{y}}{k^{2}}u(q)v(k-q)]d^{2}kd^{2}q$$

\noindent and the exponent is expressed simply as:

$$e^{\eta_{3}\Delta v +\eta_{2}\Delta u}=
Exp[\eta_{3}\int (1-e^{iQ_{x}r})v(Q)d^{2}Q +
\eta_{2}\int (1-e^{iQ_{x}r})u(Q)d^{2}Q]$$

\noindent 
It will be come clear below that transverse velocity differences $\Delta v$
obey gaussian statistics and the longitudinal ones $\Delta u$ 
are very close to
gaussian. Then, substituting the above 
 expressions into $I_{p}$ and expanding 
the exponent 
we generate an infinite series involving various products  of $u(q)$'s and 
$v(q)$'s. In case of an incompressible, statistically isotropic gaussian 
velocity field, we are dealing with, these products are split
into pairs:

$$<v_{i}(q)v_{j}(Q)>\propto
q^{-\frac{8}{3}}(\delta_{ij}-\frac{q_{i}q_{j}}{q^{2}})\delta(q+Q)$$

\noindent The $k_{y}$ integration is carried over the interval
$-\infty<k<\infty$ and in the isotropic case we are dealing with the only
non-zero terms are those involving even powers of $k_{y}$. These terms are
generated by the expansion of 

$$e^{\eta_{2} \Delta u}$$

\noindent They however, , being $O(\eta_{2})$, 
disappear in the limit $\eta_{2}\rightarrow 0$. 
Thus:

$$I_{p}=\eta_{3}\int d^{2}kd^{2}q
k_{y}(1-e^{ik_{x}r})\frac{k_{x}k_{y}}{k^{2}}<u(q)v(k-q)
Exp(\eta_{3}\int (1-e^{iQ_{x}r})v(Q)d^{2}Q +
\eta_{2}\int (1-e^{iQ_{x}r})u(Q)d^{2}Q)>
$$

\noindent where the $O(\eta_{2})$ contribution to the exponent is temporarily
kept to make the transformation

$$\Delta u e^{\eta_{2}\Delta u}=\frac{\partial e^{\eta_{2}\Delta
u}}{\partial \eta_{2}}$$

\noindent to (46) possible. Only after that we set $\eta_{2}=0$.
This proves that the only contribution to the 
equation for the probability density function comes from the $O(\Delta u
\Delta v)$ mixing components,   involved in  the pressure gradients. This 
relation justifies the estimate (46).

\section{Longitudinal Velocity Differences}

\noindent The remarkable fact that in the limit $\eta_{2}\rightarrow 0$ 
all contributions to the equation (25) contain 
$\frac{\partial}{\partial \eta_{2}}$ enables separation of variables:
integrating the resulting equation over $\eta_{2}$ gives the  closed equation
for $Z_{3}(\eta_{3})$. The corresponding dynamic 
 equation is linear,  meaning that transverse
velocity fluctuations do not directly contribute to the energy transfer
between different scales. This effect is possible only in 2d where the 
 $O((d-2)\frac{\partial}{\partial \eta_{3}})$ 
enstrophy production term in (22), not containing
$\frac{\partial}{\partial \eta_{2}}$,  is equal to zero.  This simplification,
combined with locality of the pressure-gradient 
 effects, allowed us to
derive a closed-form expression for $Z_{3}$. 

The role of pressure in the dynamics of tranverse components of velocity field 
is mainly restricted to control of the ``energy redistribution'' neccessary
 for generation of  
isotropic and incompressible velocity field. The longitudinal field dynamics
are much more involved. The advection  (pressure excluding) part of 
non-linearity tends to produce large
gradients of  velocity field (``shock generation''  
using the Burgers equation
phenomenology),  manifesting itself in creation of a 
constant energy flux in the wave-number space.   
Pressure is the only factor preventing the shock
formation.

Interested in the longitudinal
correlation functions we set $\eta_{3}=0$. Then, the term in (25)

\begin{equation}
\frac{\eta_{2}}{r}\frac{\partial^{2}Z}{\partial  \eta_{3}^{2}}=
\frac{\eta_{2}}{r}<(\Delta v)^{2}e^{\eta_{2}\Delta u}>\approx 
\frac{\eta_{2}A_{2}^{t}}{r^{\frac{1}{3}}}Z_{2}+O(\eta_{2}^{2};~\eta_{3}^{2};
~\eta_{2}^{2}\eta_{3})
\end{equation}

\noindent The last relation is accurate since substituting this into (25),
differentiating once over $\eta_{2}$ and setting 
both $\eta_{3}=\eta_{2}=0$ gives:

\begin{equation}
\frac{1}{r}\frac{\partial}{\partial r}rS_{2}-\frac{A_{2}^{t}}{r^\frac{1}{3}}=
\frac{\partial I_{p}(0,0)}{\partial \eta_{2}}
\end{equation}

\noindent Since $S_{2}(r)=A_{2}r^{\frac{2}{3}}$ this equation gives:
\begin{equation}
\frac{5}{3}A_{2}-A_{2}^{t}=
r^{\frac{1}{3}}\frac{\partial I_{p}(0,0)}{\partial \eta_{2}}
\end{equation}

\noindent which, according to (12)  is exact since 
$\frac{\partial I_{p}(0,0)}{\partial \eta_{2}}=0$ (see below). 

Let us consider some general properties of the pressure term $I_{p}$ in the
limit $\eta_{3}\rightarrow 0$. We have:

\begin{equation}
I_{p}\approx \eta_{2}<(\frac{\partial p(2)}{\partial x_{2}}-\frac{\partial
p(1)}{\partial x_{1}})
Exp(\eta_{2}\Delta u +\eta_{3}\Delta v)>
\end{equation}

\noindent 
Expanding the exponent  and recalling that the isotropic and
incompressible turbulence $\overline{\Delta u}=\overline{\Delta v}=0$ and 
$\overline{p(x)v_{i}(x')}=0$, we conclude that 

\begin{equation}
I_{p}\approx \eta_{2}<(\frac{\partial p(2)}{\partial x_{2}}-\frac{\partial
p(1)}{\partial x_{1}})
(\eta_{2}\Delta u +\eta_{3}\Delta v)^{2}+...>=O(\alpha \eta_{2}^{3}+\beta
\eta_{2}^{2}\eta_{3}+...)
\end{equation}

\noindent It is clear that the relation (48), derived above for the case of
gaussian statistics, satisfied this
general property of the flow. Thus when $\eta_{3}\rightarrow 0$,  
we approximate 

\begin{equation}
I_{p}\approx c\eta_{2}^{3}Z+G
\end{equation}

\noindent where $c$ is a yet undetermined constant and $G$ denotes the
contributions to $I_{p}$, properly modifying numerical coefficients in the 
equation (25). The presence of the $O(\eta_{2}^{3})$ distinguishes this
equation from the one for transverse PDF considered in the previous section. 
There the assumed role of pressure was limited to the mixing of various
components of velocity field. That is why all we accounted for was $O(
\Delta v \Delta u)$ contributions to pressure. Here, in addition we
also consider  $O(\eta_{2}^{3})$ contributions, responsible for prevention of
the shock formation. 
The resulting equation  is:

\begin{equation}
\frac{1}{r^{3}}\frac{\partial^{2}}{\partial \eta_{2}\partial r}r^{3}Z_{2}-
\frac{11}{5r^{\frac{1}{3}}}A_{2}^{t}\eta_{2}Z_{2}-3\eta_{2}^{2}Z_{2}-
c\eta_{2}^{3}Z_{2}=0
\end{equation}

\noindent The Laplace transform of gives equation for the probability density
$P(\Delta u,r)$:

\begin{equation}
cP_{UUU}-3P_{UU}+\frac{1}{r^{3}}\frac{\partial}{\partial r}r^{3}UP+
\frac{11 A_{2}^{t}}{5}P_{U}=0
\end{equation}
\noindent Seeking solution in a scaling form (the parameter $c$ will be
determined below)

\begin{equation}
P(U,r)=\frac{1}{r^{\frac{1}{3}}}F(\frac{U}{r^{\frac{1}{3}}})
\end{equation}

\noindent we obtain
\begin{equation}
cF_{xxx}-3F_{xx}+(b-\frac{x^{2}}{3})F_{x}+\frac{8}{3}xF=0
\end{equation}
 
\noindent
Where $b=\frac{11}{3}A_{2}$. 
All,  but one, 
term  in (65)  change 
sign wnen $x\rightarrow -x$. The  $O(F_{xx})$ symmetry-breaking contribution 
 is neccessary for existence of the non-zero energy flux. 
Assuming for a time being that, in accord with numerical and physical
experiments, that the flux is small (see relation (2)), we first neglect the
$O(F_{xx})$- contribution, find solution and then take it into account
perturbatively. The equation is:

\begin{equation}
cF^{o}_{xxx}+(b-\frac{x^{2}}{3})F^{o}_{x}+\frac{8}{3}xF^{o}=0
\end{equation}

\noindent with solution:

\begin{equation}
F^{o}=e^{\frac{x^{2}}{2A_{2}}}
\end{equation}

\noindent where $c=\frac{A_{2}^{2}}{3}$. 
If $A_{2}>>1$, then the neglected $F_{xx}=O(1/A_{2})$ term is small. 
This means
that the odd-order moments, computed with the PDF,  which is a solution of
(65),  must be small in a sense defined by the relation (2).  At the same
time the even-order moments must be close to the gaussian ones.

Analytic solution of (65) is difficult. However, one can evaluate 
all moments $\frac{S_{n}}{r^{\frac{n}{3}}}=A_{n}$ in terms of only one
parameter $A_{2}$:

\begin{equation}
S_{n+1}=-\frac{3}{n+10}(-\frac{A_{2}^{2}}{3}n(n-1)(n-2)S_{n-3}-3n(n-1)S_{n-2}-\frac{11}{3}A_{2}nS_{n-1})
\end{equation}

This relation gives: $A_{1}=0; A_{3}=3/2; A_{4}=3; A_{5}=12.43A_{2}; A_{6}=
15A_{2}^{3}-36; A_{7}=37.71A_{4}$ etc. These  numbers can be tested in
numerical experiments.  The one-loop renormalized perturbation expansions give 
$A_{2}\approx 10$, while numerical simulations are consistent with
$A_{2}\approx 12$. Keeping these numbers in mind, it follows from (68)
that the accurate measurements of the odd-order moments is the 
only way to verify
predictions of the present theory. The deviations of the 
even-order moments from the gaussian ones are too small to be detected by both
physical and numerical experiments. It can be checked that the ratios

$$s_{2n+1}=\frac{S_{2n+1}}{S_{2n}^{\frac{2n+1}{2n}}}$$

\noindent vary in the interval $0.04-0.1$ for $2<n<10$ and $A_{2}\approx 10$. 
With $A_{2}\approx 12$ these numbers decrease even more.

\section{Summary and Conclusions}

\noindent The experimentally observed gaussian or very close to it statistics
of transverse velocity differences was   extremely puzzling  since, on
the first glance,
it is incompatible with the non-trivial Kolmogorov scaling, 
resulting from strong non-linearity of the problem.  The most surpring and
interesting result, derived in this paper, is that due to the symmetries of
the problem the equation, governing probability density function of transverse
velocity differences, has one derivative less than the one corresponding to
the longitudinal differences. This means, in turn, that  transverse
components of 
velocity field are governed by a non-local linear, equation, driven by a
universal,  non-local, solution-depending gaussian force. This reduction,
resembling the super-symmetry effects in field theory, is surprising if not
miraculous. 
The non-local equation in the physical space, obtained above,
corresponds to the Langevin equation in the Fourier space:

\begin{equation}
v_{t}(k)+c_{\nu}P^{\frac{1}{3}}k^{\frac{2}{3}}v=f_{R}(k,t)
\end{equation}

\noindent where $c_{\nu}$ is an amplitude of 
``effective'' (turbulent) viscosity and 

\begin{equation}
\overline{f_{R}(k,t)f_{R}(k',t')}\propto k^{-1}\delta(k+k')\delta(t-t')
\end{equation}

\noindent used in [9]-[10] 
in the renormalization group treatments  of fluid turbulence.

\noindent 
The irrelevance of the 
dissipation terms in two-dimensional turbulence 
makes the problem much more tractable than its
three-dimensional counterpart. Still, in order to close the equations 
for  probability density of velocity field
one needs an expression for the pressure contributions. 
The situation  is even more simplified by the fact
that the large-scale-dominated single-point variables are time-dependent and
must decouple from the steady-state small-scale dynamics. That is why one can
use an assumption about locality of the pressure gradient effects leaving only
the mixing $O(\Delta u \Delta v)$ contributions to the two-point pressure
difference. It can be tested by a mere accounting that all other contributions
to the expression for $I_{p}$  involve one or more $U_{+}$'s and leading to
the time-dependent result. This means that they must
disappear from the steady state equations (25) and (45).
The range of
possible 
models for pressure is narrowed by a few dynamic and kinematic constraints and
by the fact that the resulting equation must give positive solution. A simple
calculation shows that the model for the pressure gradient terms, introduced
in this paper, is consistent 
with the derived gaussian statistics.

The equations for  PDF of longitudinal velocity differences do not
correspond to  linear dynamics. Still, the derived solution only slightly 
deviates from
gaussian. This is possible due to the relative smallness of the energy flux 
in two dimensions. 

The results presented here seem to agree with both physical and numerical
experiments. The obtained close-to-gaussian statistics justifies various
one-loop renormalized perturbation expansions giving $A_{2}\approx
10-12$. Using this number we realize that it is extremely difficult to 
experimentally detect deviations from the gaussian statistics. 
Still, some fine details of the present theory, related to the pressure 
gradient-velocity correlation functions can be tested numerically. 
In addition, measurements of a few odd-order moments can shed some light on
validity of the present theory.

The equations and solution presented here leave one question unanswered: 
are these 
${\bf the}$ solutions or not? Our experience with the Burgers 
 and 2d Navier-Stokes
equations teach us that it is very difficult to find a self-consistent closure 
leading to the positive solution for the PDF's. Stretching this statement 
a bit 
we feel that a closure, satisfying dynamic constraints and leading 
to a a plausable
solution has a great chances to be correct. 

\noindent Absense of intermittency in a  steady-state 
developing inertial range discovered 
in two-dimensional turbulence [2]-[4] seems to be a general
phenomenon observed in a drift-wave turbulence [11] and in a one-dimensional
model of a passive scalar advected by a compressible velocity field [12].
These observations support our understanding of intermittency as a phenomenon 
originating  from
interaction of the large and small-scale velocity fluctuations. In a
developing statistically steady 
inertial range,  were the integral scale is strongly time-dependent,
these interactions must be small for the small-scale steady state to exist. 
At the later stages the finite size effects, destroying   
time-independence of the small scale
dynamics,  lead to formation of coherent structures and new dynamic phenomena
which are beyond the scope of the present theory.

\section{Acknowledment} I am grateful to A. Polyakov, M. Vergassola, 
M.Chertkov, B.Shraiman, Y. Sinai  and I. Kolokolov for many 
interesting and
illuminating discussions.

\noindent {\bf references}
\\
1. R.H.Kraichnan, Phys.Fluids. {\bf 10}, 1417 (1967),
\\
2. L.M. Smith and V. Yakhot, Phys.Rev.Lett. {\bf 71}, 352 (1993)
\\
3. L.Smith and V. Yakhot, J. Fluid. Mech. {\bf 274},  115 (1994)
\\
4. P. Tabeling and J. Paret, Phys. Fluids, {\bf 12}, 3126 (1998)
\\
5.  L.D.Landau and E.M. Lifshitz, Fluid Mechanics, Pergamon Press, Oxford, 198
\\
6. A.S.Monin and A.M.Yaglom, ``Statistical Fluid Mechanics'' vol. 1, MIT Press,
Cambridge, MA (1971)
\\
7 U. Frisch, ``Turbulence'', Cambridge University Press, 1995
\\
8 .  A.M. Polyakov, Phys.Rev. E, {\bf 52}, 6183 (1995)
Phys.Rev. E {\bf 52}, 6183 (1995)
\\
9. de Dominicis and P.C.Martin, Phys.Rev.A {\bf 19}, 419 (1979)
\\
10. V. Yakhot and S.A.Orszag, Phys.Rev.Lett. {\bf 57}, 1722 (1986)
\\
11. N. Kukharkin, S.A.Orszag and V.Yakhot, Phys.Rev.Lett., {\bf 75}, 2486
    (1995)
\\
12. K. Gawedzki and M.Vergassola, COND-MAT/9811399, (1998)
\\

\end{document}